\documentclass{article}

\usepackage{arxiv}

\usepackage[utf8]{inputenc} 
\usepackage[T1]{fontenc} 
\usepackage{hyperref} 
\usepackage{url} 
\usepackage{booktabs} 
\usepackage{amsfonts} 
\usepackage{amssymb,amsmath} 
\usepackage{nicefrac} 
\usepackage{microtype} 
\usepackage{xcolor}
\usepackage{graphicx}
\usepackage{float}
\usepackage{doi}
\usepackage{siunitx}
\usepackage{multirow}
\usepackage{xcolor}

\usepackage{cite}

\title{Topological characterization of rearrangements in amorphous solids}

\author{
	\href{0000-0001-7257-8125}{\includegraphics[scale=0.06]{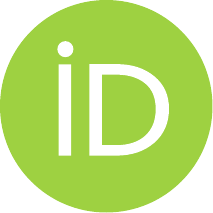}\hspace{1mm}Paul Desmarchelier} \\
	Department of Material Sciences and Engineering, Johns Hopkins University, Baltimore, Maryland 21218, USA\\
	\texttt{paul.desma@gmail.com} 
	\And
	Spencer Fajardo \\
	{	Department of Material Sciences and Engineering, Johns Hopkins University, Baltimore, Maryland 21218, USA}
	\And
	\href{0000-0002-8383-4259}{\includegraphics[scale=0.06]{orcid.pdf}\hspace{1mm} Michael  L. Falk} \\
	{	Department of Material Sciences and Engineering, Johns Hopkins University, Baltimore, Maryland 21218, USA}\\
	{Department of Mechanical Engineering, Johns Hopkins University, Baltimore, MD 21218, USA }\\
	{Department of Physics and Astronomy, Johns Hopkins University, Baltimore, MD 21218, USA}\\
	{Hopkins Extreme Materials Institute, Johns Hopkins University, Baltimore, MD 21218, USA}
}

\renewcommand{\headeright}{A Preprint}
\renewcommand{\shorttitle}{Topological characterization of rearrangements in amorphous solids }

\hypersetup{
pdftitle={},
pdfsubject={cond-mat, nanomaterials},
pdfauthor={Paul Desmarchelier },
pdfkeywords={Draft},
}

\begin{document}
\maketitle

\begin{abstract}
	In amorphous materials, plasticity is localized and occurs as shear transformations. It was recently shown by Wu et al. that these shear transformations can be predicted by applying topological defect concepts developed for liquid crystals to an analysis of vibrational eigenmodes [Wu et al.; Nat. Com.,2023]. This study relates the -1 topological defects to the displacement fields expected of an Eshelby inclusion, which are characterized by an orientation and the magnitude of the eigenstrain. A corresponding orientation and magnitude can be defined for each defect using the local displacement field around each defect. These parameters characterize the plastic stress relaxation associated with the local structural rearrangement and can be extracted using the fit to either the global displacement field or the local field. Both methods provide a reasonable estimation of the MD-measured stress drop, confirming the localized nature of the displacements that control both long-range deformation and stress relaxation.
	
\end{abstract}

It has been long appreciated that structural defects play an essential role in the mechanical behavior of crystalline materials. For instance, their yield stress is determined by the behavior of dislocations \cite{vitek_structure_1992}. 
These defects are described as breaks in the invariance of the crystal lattice \cite{vitek_atomic_2011}. 

In amorphous materials, there is no lattice structure, and it is challenging to define discrete defects linked to deformation.
The lack of clearly defined defects inhibits the development of a deformation theory for amorphous materials that is specifically linked to aspects of the atomic structure~\cite{falk_deformation_2011,tanguy_elasto-plastic_2021}. Consequently, predicting structural features that give rise to a larger-scale deformation remains an area of active investigation \cite{richard_predicting_2020}. 

The sites, where the structural relaxation takes place, are characterized by large non-affine displacements \cite{falk_dynamics_1998} and high potential energy release \cite{fusco_role_2010}. These shear transformations (STs) are irreversible atomic displacements that contribute to the transition of the system from one inherent structure to another \cite{stanifer_avalanche_2022}. STs can be quantified in terms of number \cite{delogu_id_2007,yu_atomi_2021} and activation energy \cite{xu_strain-dependent_2017,boioli_shear_2017}. As such they have been integrated as a fundamental micromechanism into numerous constitutive equations \cite{bouchbinder_athermal_2007,manning_strain_2007,rycroft_eulerian_2015,hinkle_coarse_2017,kontolati_manifold_2021}.

Previous studies have characterized the spatial rearrangement taking place in the STs in 2D \cite{tanguy_plastic_2006,cao_strain-rate_2013,jin_using_2021} and 3D \cite{zink_plastic_2006} simulated glasses, and experimentally in colloidal \cite{schall_structural_2007} and metallic \cite{ma_nano_2015,kang_direct_2023} glasses. They are typically analyzed in terms of the non-affine displacements occurring during a structural relaxation, a review of which has been written by Nicolas and Röttler \cite{nicolas_orientation_2018}. In particular, these STs can be described as quadrupolar zones \cite{sopu_atomic-level_2017}. Moreover, an orientation can be assigned to the quadrupolar zones \cite{nicolas_orientation_2018}. More recently, the concept of topological defect (TD) from the liquid crystal literature has been applied to the displacement field to charcterize STs \cite{baggioli_plasticity_2021}.
Another ongoing and important effort is being made to improve the predictions of the positions of sites susceptible to local rearrangements: the shear transformation zones. Many indicators have been developed, an overview can be found in the review paper of Richard et al. \cite{richard_predicting_2020}. One of the approaches is to consider the eigenmodes of the system, with features of the lowest frequency eigenmodes predicting the next instability \cite{mazzacurati_low_1996,tanguy_vibrational_2010}. Very recently, Wu et al. also identified TDs within low frequencies eigenmodes to predict the STs \cite{wu_topology_2023}.

%

These small-scale STs have long-range repercussions. In particular, the quadrupolar zones gjve rise to long-range elastic deformations, and these can be described using the Eshelby inclusion model \cite{eshelby_determination_1957}. This has been extensively incorporated into mesoscale models of glass plasticity \cite{tanguy_plastic_2006,dasgupta_yield_2013,castellanos_history_2022}. Notably, Albaret et al. showed that the stress drop due to relaxation can be accurately estimated using the position and characteristics of the Eshelby inclusions fitted locally~\cite{albaret_mapping_2016}. A similar analysis has been performed to study the orientation \cite{nicolas_orientation_2018}.

In this paper, we show that the plastic relaxation can be characterized using the core of the STs and their immediate surroundings. First, the STs are located by characterizing the topological defects within the displacement field, and measuring the orientation and a magnitude of each defect. We then show that this information can be used to reproduce the entire displacement field using the Eshelby inclusion model. Finally, we relate the orientation and magnitude of the STs to the stress drops using values estimated from either the global displacement field or from the local displacement around the STs only. 


The glass samples studied here are 2D binary Lennard-Jones glass squares with a side length of 98.8 (Lennard-Jones reduced units) and containing 10,000 atoms, created using the same potential parameters and the slow quench approach described by Barbot et al. \cite{barbot_local_2018}. Fourteen independent samples using different initial spatial distributions are created.
These glasses are then deformed in simple shear using an Athermal Quasi Static (AQS) algorithm. That is, the system is deformed step-wise with a strain step ($\delta\gamma$) of \num{1e-5} and then relaxed to mechanical equilibrium using a conjugate gradient method before performing the next strain step. This is repeated until a strain ($\gamma$) of 0.5 is reached. Lees-Edwards periodic boundary conditions are maintained throughout, and all simulations are performed using LAMMPS \cite{LAMMPS}.

\begin{figure}[h]
	
	\includegraphics[width=\linewidth]{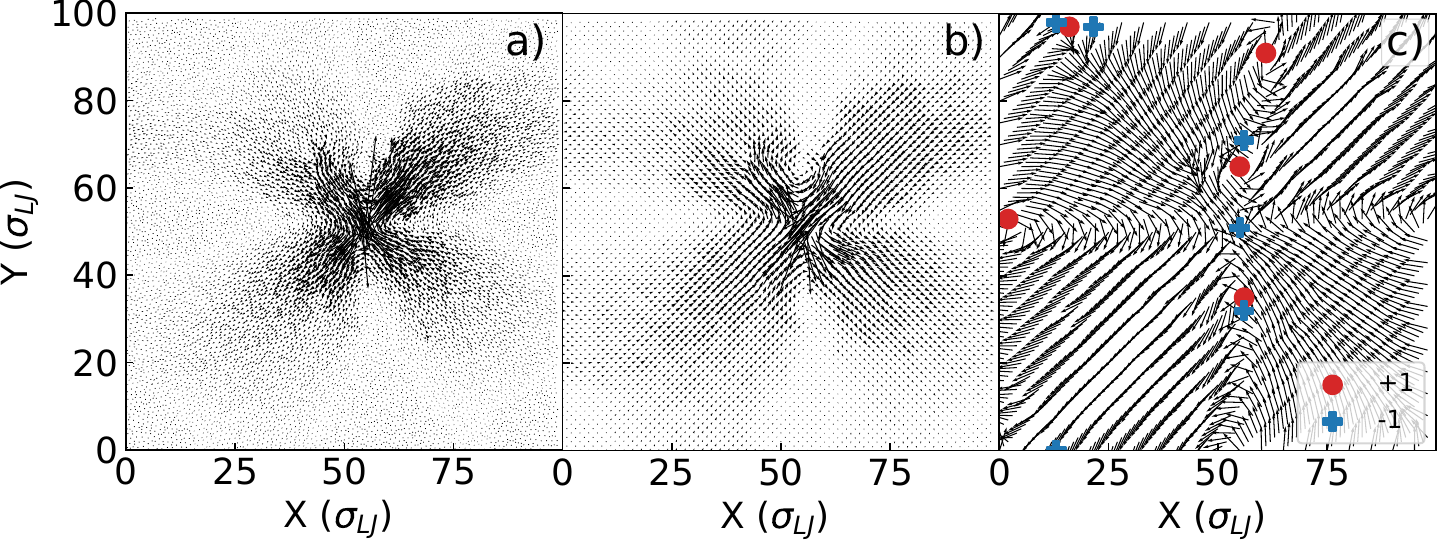}
	\caption{	\label{fig:DisToDe}Representation of the non-affine displacements during a plastic event: a) displacements mapped on atomic positions b) coarse-grained displacements c) normed coarsed-grained displacement vectors (black arrows) used to determine the topological defect positions (blue crosses for -1 defects and red circles for +1 defects )}
\end{figure}

During the deformations, the stress and strain of the whole box are recorded at each deformation step, and the atomic positions are recorded before and after each plastic event. A plastic event, or stress relaxation event, is defined in this study as a shear step that results in the global stresses in the shear direction $\sigma_{xy}$ decreasing. The impact of this choice is discussed in the supplementary materials. The non-affine displacements are computed using the atom positions before the event $r_{i-1}$ and after the event $r_i$: 
\begin{equation}\label{eq:NA}
	u_i=r_{i}-r_{i-1}-u_i^{aff}
\end{equation}
with $u_i^{aff}$ the imposed affine displacement ($u_x=\delta\gamma r_y,u_y=0$). To represent the non-affine displacement field corresponding to the structural relaxation, the displacements computed via equation \ref{eq:NA} are mapped onto the position of the atoms just before the stress drop with the added affine deformation, this is depicted in figure \ref{fig:DisToDe}a. Additionally, after each stress drop, the stress state after the reversion of a single strain step is computed. The occurrence of a plastic event during this step is very unlikely. This state will be useful for the computation of the shear modulus of the inherent state and is referred to as reverted state in the following.

The first step of the analysis is to identify the positions of the STs that give rise to the stress drop.
This is done by identifying topological defects in the displacement field, in a very similar way to the approach developed by Wu et al. \cite{wu_topology_2023}. First, atomic displacements are projected onto a 100 by 100 orthogonal regular grid (see figure \ref{fig:DisToDe}b) with the coarse graining function described by Albaret et al. \cite{albaret_mapping_2016} using a length of 1.17 Lennard-Jones distance units ($\sigma_{LJ}$). The topological defects are defined by the smallest closed loop for which the topological charge $q$ takes a non-zero value, as defined by the equation \cite{selinger2016introduction}
\begin{equation}
	\label{eq:def}
	\oint d\theta= 2\pi q.
\end{equation} 
Here, $\theta$ is the orientation of the displacement (as represented by the orientation of the normed vector in figure \ref{fig:DisToDe}c). As a simplification, the topological defects are computed for each point of the grid by considering a 4 by 4 square loop around the point considered. An example of the resulting charges is given in figure \ref{fig:DisToDe}c. The final position of the defect is the center of mass of the contiguous patches sharing the same topological charge. The quadrupoles and vortices arising upon deformation of a 2D glass \cite{sopu_stz-vortex_2023} will appear as -1 and +1 defects, respectively. Importantly, those are the only topological charge values that we observe, meaning that the displacement field can be described as a superposition of quadrupoles and vortices. In figure \ref{fig:DisToDe}c, a -1 topological charge is detected in the central quadrupolar zone visible in panels a) and b), as well as other charges. These other charges appear in regions where the displacements are smaller. As this analysis stems from nematics, it considers directors and not vectors; as such, the quadrupoles are four-fold symmetric.

More information about the -1 defects can be extracted from the displacement field in their immediate proximity. We are able to calculate the orientation, $\phi_{esh-loc}$ estimated using the phase shift of the inner product of the displacement of each atom and their position relative to the center of the ST, i.e.

\begin{equation}\label{eq:ang}
	\phi_{esh-loc}=\arg\left(\sum_{i=0}^{N_{shell}}\mathbf{u}_i\cdot(\mathbf{r}_i-\mathbf{x}_{esh})\exp(-j2\Omega)\right).
\end{equation}

Here, $\Omega$ is the angular position of atom $i$ relative to the Eshelby inclusion position ($\mathbf{x}_{esh}$), and $j$ is the imaginary number. The sum of equation \ref{eq:ang} runs on all the atoms within 4 interatomic distances of the center of the -1 defect considered ($N_{shell}$). The phase shift is deduced from the argument of the second term of the complex Fourier series. This approach is similar to the method \textit{MD/azi} introduced by Nicolas and Röttler \cite{nicolas_orientation_2018}.
The displacement amplitude near the defects can be used to describe the importance of the defect, both relative to other defects and relative to the global stress relaxation. This can be described using the average atomic non-affine displacement in the contiguous patch sharing the same topological charge (-1) $\langle|\mathbf{u}_{na}|\rangle_{defect}$. 
\begin{figure}[h]
	
	\includegraphics[width=.5\linewidth]{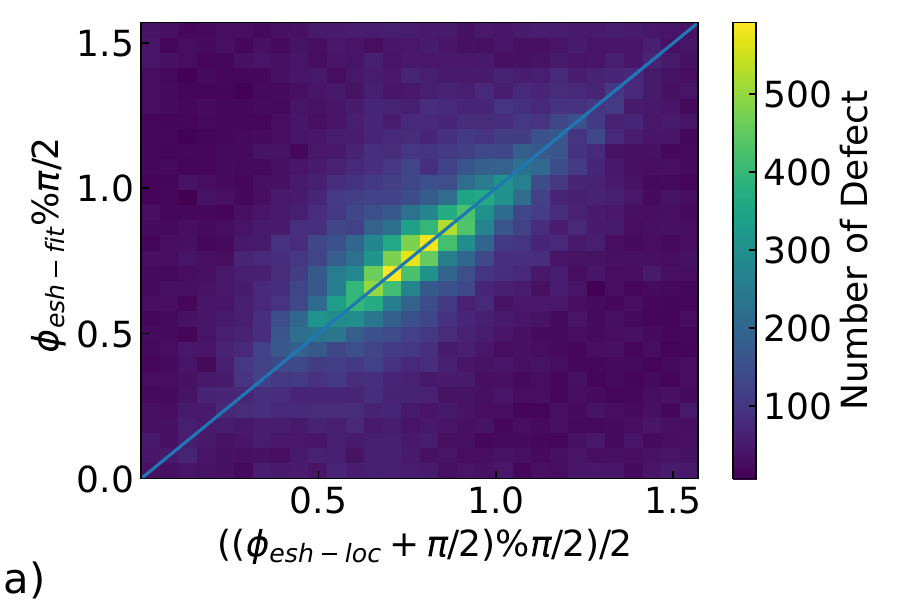}
	\includegraphics[width=.5\linewidth]{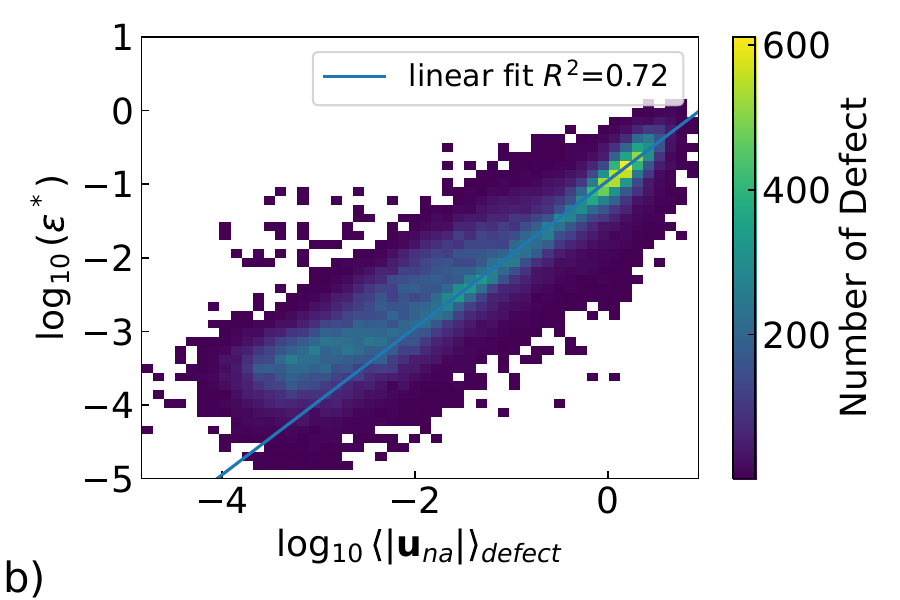}
	\caption{	\label{fig:Corr}a) Distribution of the angle $\phi$ obtained from the fitting of the whole displacement field to equation \ref{eq:disp} vis-à-vis $\phi_{esh-loc}$ the angle of the local displacement field from equation \ref{eq:ang}. The diagonal blue line represents one-to-one correspondence. b) The eigenstrain of the Eshelby equation $\varepsilon^*$ vis-à-vis the average displacement around the defect. The blue line represents the linear fit.}
\end{figure}

The -1 defects, being essentially quadrupoles, can be interpreted as the center of an Eshelby inclusion, and as such, the resulting displacement field may be computed using the following equations, derived by Dasgupta et al. in 2D \cite{dasgupta_yield_2013},
\begin{equation}	\label{eq:disp}
	\begin{split}
		u_x&=\frac{\varepsilon^*}{4(1-\nu)}\frac{a^2}{r^2}\left\{\left[ 2(1-2\nu)+\frac{a^2}{r^2} \right] \left[ x\cos2\phi+y\sin2\phi \right]+\left[1-\frac{a^2}{r^2} \right]\left[ \frac{(x^2-y^2)\cos2\phi+2xy\sin2\phi}{r^2} \right]2x \right\},
		\\
		u_y&=\frac{\varepsilon^*}{4(1-\nu)}\frac{a^2}{r^2}\left\{\left[ 2(1-2\nu)+\frac{a^2}{r^2} \right] \left[ x\sin2\phi-y\cos2\phi \right]+\left[1-\frac{a^2}{r^2} \right]\left[ \frac{(x^2-y^2)\cos2\phi+2xy\sin2\phi}{r^2} \right]2y \right\}.
	\end{split} 
\end{equation}
Here $a$ isthe radius of the inclusion; $\phi$ is the orientation of the quadrupolar zone; $r$ is the distance to the center of the core; $x$ and $y$ describe the position of the center of the inclusion; $\varepsilon^*$ is the eigenstrain magnitude, and $\nu$ the Poisson ratio. The Poisson ratio is set at 0.46, based on an AQS tensile deformation simulation of the same glass, and is considered to be homogeneous and constant through the simulation. It is also important to note that here we assume a traceless eigenstrain as Dasgupta et al. \cite{dasgupta_yield_2013}, that is, that locally the Eshelby inclusion undergoes a transformation at constant volume.
Equation \ref{eq:disp} provides a solution for the displacements outside the core ($r>a$) in an infinite homogeneous medium and can be used to fit the displacement field obtained via MD. To this end, the global field is reproduced by summing the displacement of each -1 defect obtained by applying equation \ref{eq:disp}.
The parameter $a$ is set to 2 $\sigma_{LJ}$, and $\varepsilon^*$ and $\phi$ are fitted using a conjugated direction method. As a result, for each event, the displacement field is fitted using two parameters ($\varepsilon^*$ and $\phi$) per defect. Importantly, only the atoms at a distance greater than $a$=2 $\sigma_{LJ}$ away from defects are considered. Details about the fitting process and visualizations of the fitted displacements can be found in the supplementary materials.

The parameters $\phi$ and $\varepsilon^*$, describe the displacements associated with the local ST while $\phi_{esh-loc}$ and $\langle|\mathbf{u}_{na}|\rangle_{defect}$ characterize the displacement field surrounding the ST. Their correlation is displayed in figure \ref{fig:Corr}. The link between $\phi_{esh-fit}$ and $\phi_{esh-loc}$ appears clearly in figure \ref{fig:Corr}a with a diagonal distribution showing a one-to-one correspondence. It spans over $[0,\pi/2]$ because of the four-fold symmetry of the quadrupolar zones. The Pearson correlation between the two parameters is 0.26. It also appears that the distribution is not uniform over the whole set of angles, but rather appears centered around $\pi/4$. This $\pi/4$ value corresponds approximately to the orientation of the rearrangement shown in figure \ref{fig:DisToDe} and aligns with simple shear in the $x$ direction \cite{nicolas_orientation_2018,sopu_stz-vortex_2023}. 

The Eshelby inclusion eigenstrain magnitude $\varepsilon^*$ can be characterized by the displacement within the core, and approximated using $\langle|\mathbf{u}_{na}|\rangle_{defect}$. As shown in figure \ref{fig:Corr}b, these terms can be linked through a linear fit with a $R^2$ of 0.72. In this figure, one can also notice that there is an important concentration of $\varepsilon^*$ values around \num{1e-1} and a long tail up to \num{1e-4}. It can be noted that less than 1 \si{\percent} of the defects have an associated $\varepsilon^*$ below  \num{1e-10} and are not displayed in figure \ref{fig:Corr}b.

\begin{figure}[h]
	\
	\includegraphics[width=.5\linewidth]{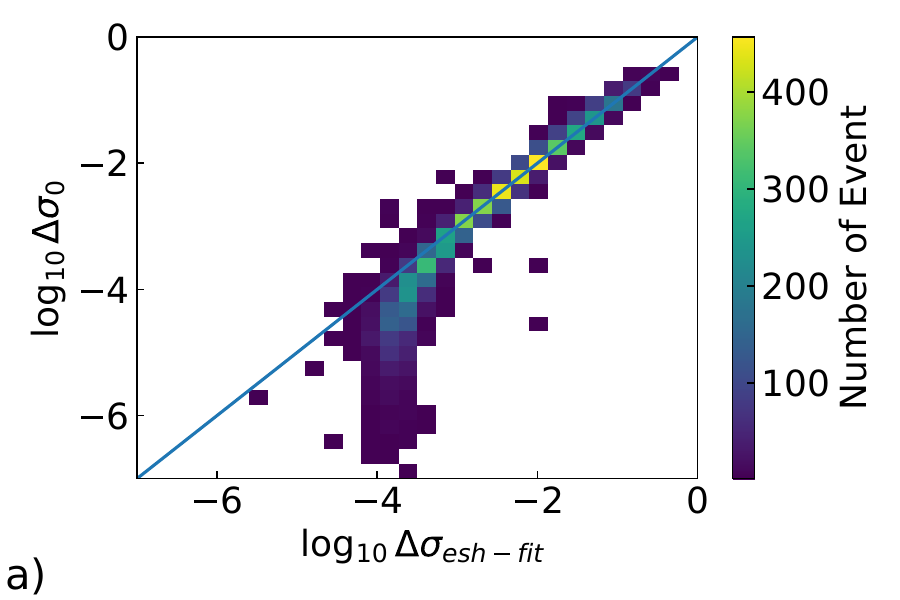}
	\includegraphics[width=.5\linewidth]{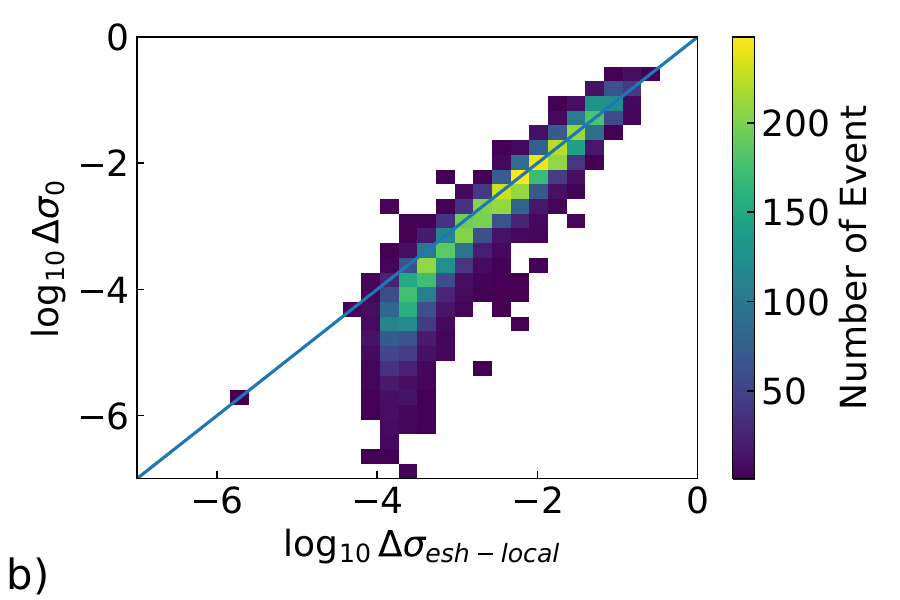}
	\caption{\label{fig:drop} 2D histogram of the stress drop of each event $\Delta \sigma_{0}$: a) as a function of the stress drop computed from equation \ref{eq:sigma} using the parameters obtained with the Eshelby fit $\Delta \sigma_{esh-fit}$, b) compared to using the defect description parameters $\Delta \sigma_{esh-loc}$. The blue lines representing where the two are exactly equal.}
\end{figure}

As described by Albaret et al. \cite{albaret_mapping_2016} the stress drop ($\Delta \sigma_{xy}$) associated with an event can be recomputed by summing the individual contribution of each inclusion,
\begin{equation}\label{eq:sigma}
	\Delta\sigma_{xy}=\sum^{n_{in}}\frac{a^2\pi}{V}G_i\varepsilon^{*}\sin(2\phi),
\end{equation}
with V the volume of the simulation cell and $G_i$ the shear modulus during event $i$.  This has been obtained by identifying the parameters of equation \ref{eq:disp} in the equation derived by Albaret et al. \cite{albaret_mapping_2016}. From this point on, unless otherwise stated,  $\Delta\sigma$ or $\sigma$
will refer to the xy component, the only one being treated in this 
study.  The shear modulus is estimated using the same method as Albaret et al.: it is the difference between the $\sigma_{xy}$ after the event and on the reverted state divided by $\delta\gamma$. This gives an estimation of the shear modulus of the inherent structure in the absence of plasticity. Other approximations not relying on the reversion of events can be used but do not perform as well for the prediction of $\Delta\sigma$ (see supplementary materials). As for equation \ref{eq:disp}, a homogeneous medium is assumed. As shown in figure \ref{fig:drop}a, the MD-derived stress drop $\Delta \sigma_{0}$ and the value derived from the displacement field fit through equation \ref{eq:sigma}, $\Delta \sigma_{esh-fit}$ have almost a one-to-one correspondence. This is particularly true for stress drops above \num{1e-3}, which account for most of the stress relaxation \cite{SM}. The Pearson correlation coefficient between fitted and MD stress drops is 0.97. 
	More importantly, using the parameters derived from the field displacement in the vicinity of the defects alone yields a similarly good correlation, with the Pearson correlation coefficient shifting to 0.92 as shown in figure \ref{fig:drop}b. It is also worth mentioning that the distribution of stress drops is not uniform over a range but that there is an important concentration around \num{1e-2} and a long tail at lower values, much like for the distribution of the eigenstrain magnitudes.

\begin{figure}[h]
		
		\centering
		\includegraphics[width=.49\linewidth]{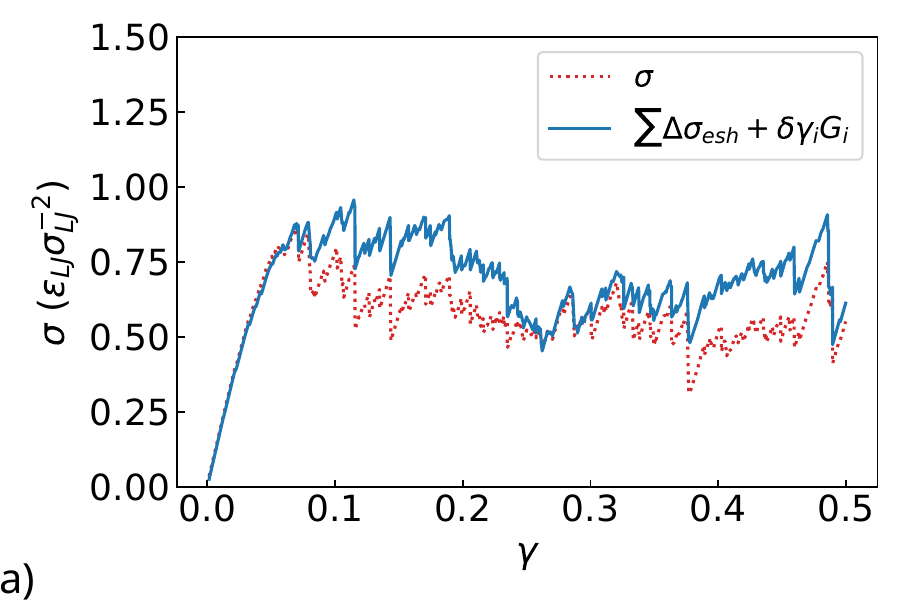}
		\includegraphics[width=.49\linewidth]{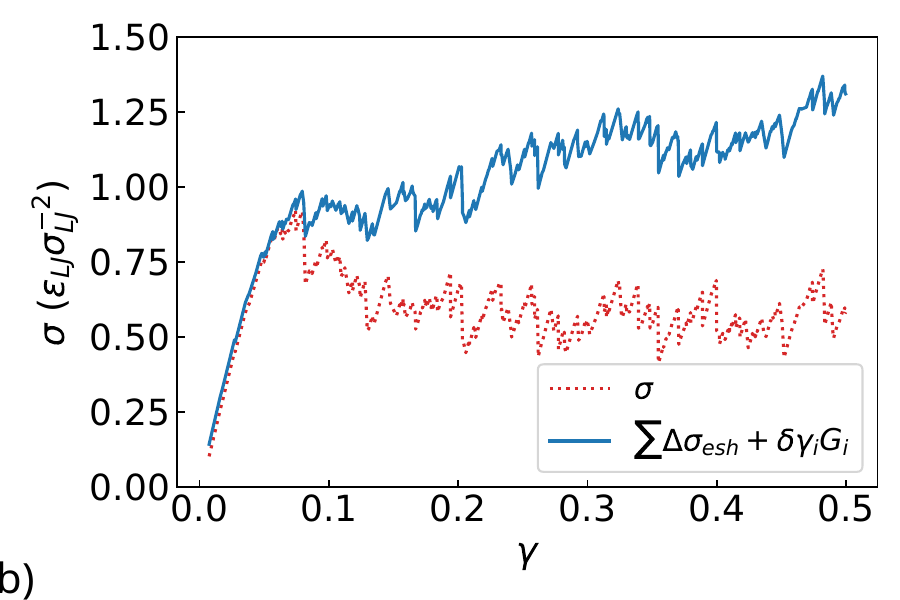}
		
		\caption{	\label{fig:stressStrain} Stress-strain curves from the simulation (dotted red line), and estimated from equation \ref{eq:sigma} with the parameters fitted from equation \ref{eq:disp},a) example with good correspondence and b) poor correspondence. }
	\end{figure}

	The stress-strain curves can be reproduced by adding the elastic part to the result equation \ref{eq:sigma},
	\begin{equation}
		\sigma^n_{esh}=\sum^{n}_i\Delta\sigma_{esh-i}+\delta \gamma_i G_i,
	\end{equation}
	and are displayed in figure \ref{fig:stressStrain}. Despite the close correspondence observed in figure \ref{fig:drop}a the stress is overestimated, meaning that the stress relaxation computed with equation \ref{eq:sigma} is underestimated. This underestimation averages at -3 \si{\percent} in relative error, the full distribution can be found in the supplementary material.

	We have shown that it is possible to infer the stress drops linked to plastic events with fair accuracy, considering either the entire displacement field or only the amplitude and orientation of the field around the -1 topological defects.

 This analysis is in essence close to the $D^2_{min}$ analysis introduced by Falk and Langer \cite{Falk1998} which is still widely used \cite{richard_predicting_2020}, both quantify heterogeneities in the non-affine displacement field. But the analysis presented here contains more information and can be used to identify quadrupolar zones (-1 defects) and vortices (+1 defects) \cite{wu_topology_2023}. As described by Sopu et al.\cite{sopu_stz-vortex_2023}, those two structures can be used to describe plastic phenomena in glasses, notably shear banding. Moreover, it is less complex than the method of Fusco et al., which relies on the spatial decay of the plastic energy \cite{fusco_role_2010}.

Albaret, et al. previously reproduced the stress-strain curves from the displacement field for a 3D amorphous silicon, with a very good accuracy \cite{albaret_mapping_2016}. This accuracy can partially be attributed to the consideration of a variable inclusion size. We and others have relied on a constant $a$ \cite{dasgupta_yield_2013}, a ST size that varies from event to event can also be estimated based on the spatial decay of plastic potential energy \cite{albaret_mapping_2016}, or the number of atoms having a high $D^2_{min}$ \cite{nicolas_orientation_2018}. In those studies, $a$ ranged from 2 to 10 interatomic distances for a-Si and 2 to 20 for the LJ glasses. However, the estimation of the ST size is bound to the estimation of the position of the ST, and this position does not always precisely match the topological defect position. It might be possible to circumvent this issue by isolating every STs by pinning the atoms of other STs in an auxiliary simulation, as proposed by Nicolas and Röttler \cite{nicolas_orientation_2018}. However, this will most probably impact $\varepsilon^*$ due to the altered boundary conditions in the immediate surroundings of the inclusion.
Moreover, using relative fitting error as defined by Albaret et al, we find a similar distribution. \cite{albaret_mapping_2016} . It is rather spread and provides an estimate with high error for some events, but achieved a lower error for the high-stress drop events (see supplementary materials).

This study relies on the formulation of the Eshelby displacement field, which assumes a homogeneous infinite medium. Thus, the solution for individual inclusion does not consider self-interaction through the periodic boundary condition, and a size effect might arise. If there is indeed a size effect observable for smaller sizes, the error seems to stabilize within 3\% from a size of 100 $\sigma_{LJ}$ on  (see supplementary materials). Moreover, the relative stress drop error distribution is not dependent on the number of events \cite{SM}. This hints that the underestimation is not caused by neglecting interaction between defects, and much less by self-interaction through the boundary conditions.
Albaret et al. considered the interaction thanks to the superposition of displacement fields due to inclusion in periodic images\cite{albaret_mapping_2016}, but in our case it does not improve the results and increases the computational cost.

We conclude that there is an essential relationship between the rearrangements that control plasticity and -1 topological defects in the displacement field. They can be used to identify the center of the shear transformations from which quadrupolar relaxation arise. An orientation and magnitude of the eigenstrain can be assigned to these centers either by fitting the displacement field using the Eshelby inclusions model or using the non-affine displacement in the vicinity of the defect. Using the characteristics of the inclusions obtained from the fits or from the local displacement field, it is possible to obtain a reasonable approximation of the stress relaxation. This reaffirms, with earlier studies~\cite{hieronymus-schmidt_shear_2017,dasgupta_yield_2013,tanguy_elasto-plastic_2021,stanifer_avalanche_2022}, that rearrangements in amorphous materials are composed of discrete, local STs that can be enumerated and characterized as such. This is likely true not only in the 3D covalent glasses previously studied by Albaret, but across a wide range of glasses, including metallic glasses and 2D systems. The topological defect concept provides an unambigous methodology for locating and characterizing such STs.

\section*{Acknowledgment}
This work supported by the US National Science Foundation (NSF) under Grant No. DMR-2323718/2323719/2323720 and was carried out at the Advanced Research Computing at Hopkins 
(ARCH) core facility (rockfish.jhu.edu), which is supported by the NSF under grant number OAC 1920103. The authors would like to thank T. Albaret, W. Kob and T. Curk for fruitful discussion.

\bibliographystyle{ieeetr}
\bibliography{FalkGroup}

\end{document}


\maketitle

\renewcommand{\thefigure}{S\arabic{figure}}

\renewcommand{\theequation}{S\arabic{figure}}

\section{Contribution to Relaxation and Event Definition}\label{app:drop}

\begin{figure}[H]
	\centering

	\includegraphics[width=.5\linewidth]{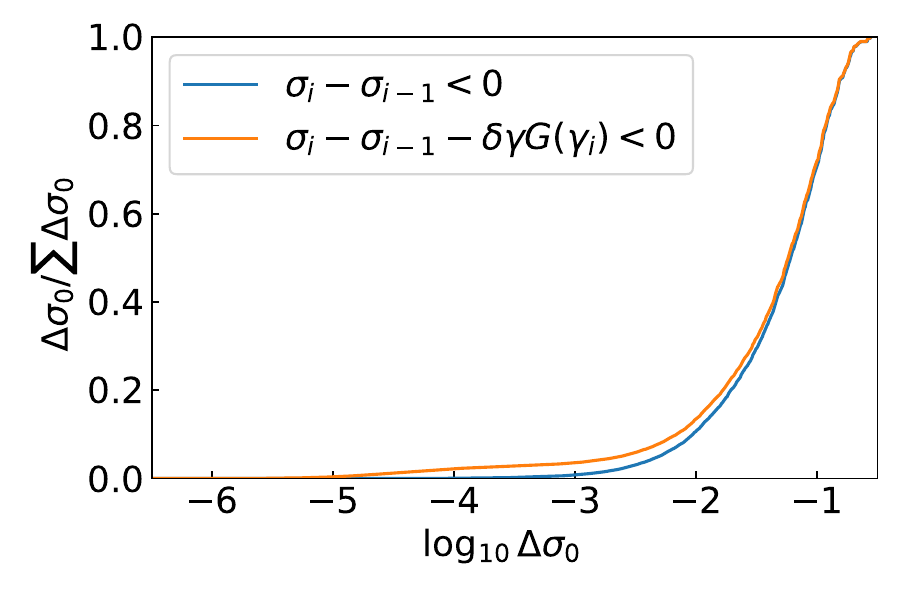}
	\caption{	\label{fig:1DHist}Normed cumulative contribution to the total stress drop as a function of stress relaxation, with (orange line) or without considering the elastic part of the stress drop (blue line). }
\end{figure}

The normed cumulative contribution to the total stress relaxation of events as a function of $\Delta \sigma_0$ is represented in figure \ref{fig:1DHist}. 
It shows the impact of the choice of event detection. In the present study, the relaxation events are defined as steps $i$ where $\sigma_i-\sigma_{i-1}<0$. However, this neglects steps where the plastic stress relaxations are lower than the elastic stress increases ($\delta\gamma G\approx$ \num{2e-4}). Those steps can be considered by using the following relation to define events: $\sigma_i-\sigma_{i-1}-\delta\gamma G(\gamma_i)<0$. In figure \ref{fig:1DHist} the two are represented, and it appears that the error made by neglecting $\delta\gamma G(\gamma_i)$ (the difference between the curves) amounts to less than 5\% of the stress relaxation. From the distribution, it also appears that most of the stress relaxation occurs for drops larger than \num{1e-3}. So that, with an accurate prediction of relaxation above this value, most of the behavior should be reproducible. This may contribute to the under estimation of the stress relaxation seen in figure 4.

\section{Fitting error}\label{app:err}
\begin{figure}[H]
	\centering
	
	\includegraphics[width=.5\linewidth]{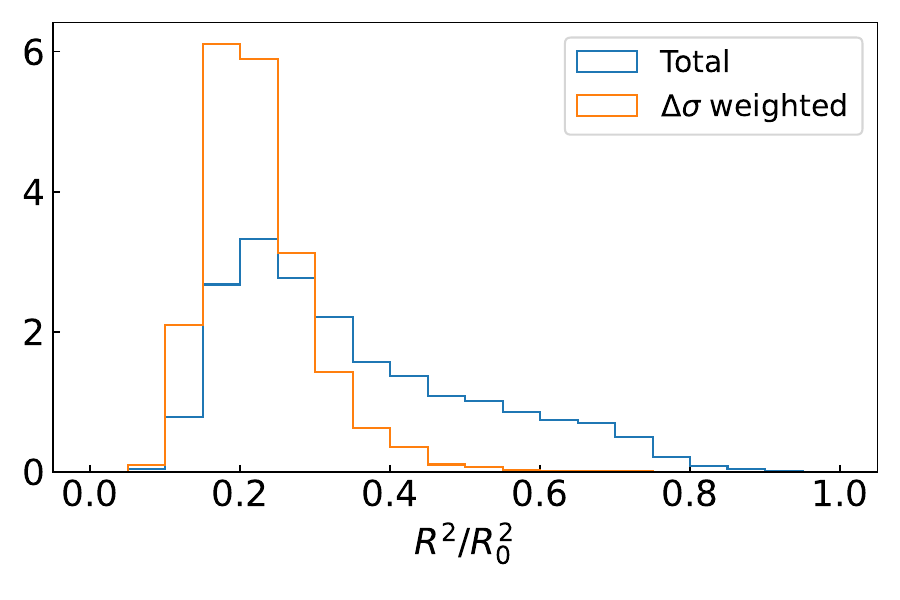}
	
	\caption{	\label{fig:err}Fitting error as defined by Albert et al. for all stress drops (orange), and for all events weighted by $\Delta \sigma$ (blue). }
\end{figure}

The function that is minimized by the fitting procedure is defined as:
\begin{equation}
	R^2=\sum^{N_{out}}\left|\mathbf{u}_{NA}-\sum^{N_{incl}}\begin{pmatrix} u_x \\ u_y \end{pmatrix}(\mathbf{r}-\mathbf{r}_{incl})\right|^2
\end{equation}
with$ \sum^{N_{out}}$ the sum over the atoms that are out of the inner core (at a distance of more than $a$ away) of all inclusions, $\mathbf{u}_{NA}$ the non-affine displacements, and $\sum^{N_{incl}}$ the sum over the displacement due to each inclusion. As suggested by Albaret et al. \cite{albaret_mapping_2016}, the final value of $R^2$ is compared to $R_0^2=\sum^{N_{out}}|\mathbf{u}_{NA}|^2$ to compute a relative error, the result is represented in figure \ref{fig:err}. The error seems to depend on $\Delta\sigma$, as can be seen by the shift in the peak of the error when weighted by the stress drop; the resulting error distribution is centered around a lower value. The displacements due to high-stress drop events are thus easier to reproduce using equation 4, centered on the main TDs. This behavior was already shown by Albaret et al. for events with large plastic energy \cite{albaret_mapping_2016}.

\section{Displacement field examples}\label{app:disp}
\begin{figure}[H]
	\centering
	
	\includegraphics[width=.5\linewidth]{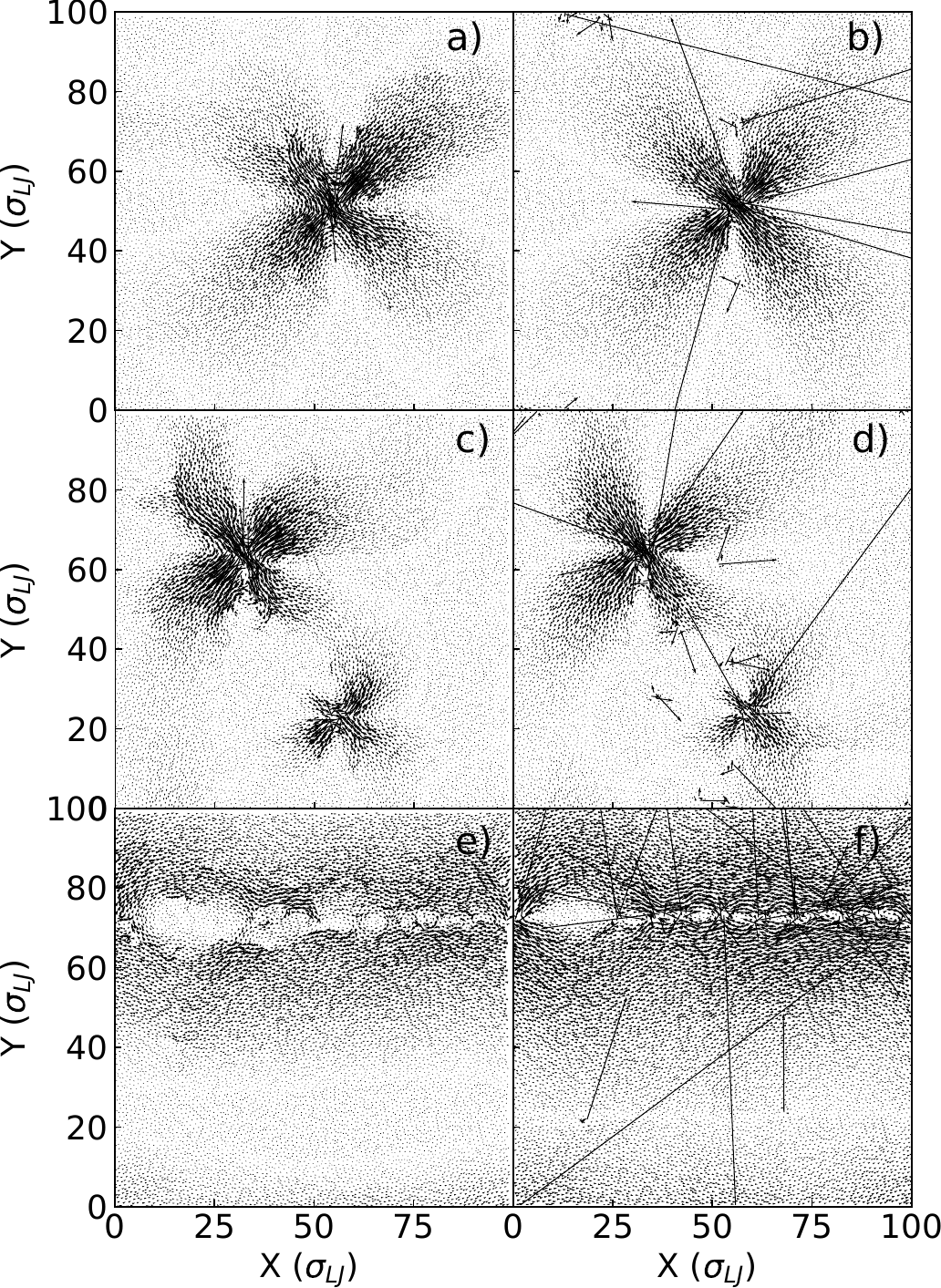}
	\caption{	\label{fig:DispRep} Representation of the non-affine displacements from the MD simulation (a, c, e) and fitted by equation 4 (b, d, f), each row representing the same event. }
\end{figure}

The displacement field obtained through the MD simulation and the one resulting from the fit of $\varepsilon^*$ and $\phi$ from equation 4 for the selected -1 defect are displayed in figure \ref{fig:DispRep}. The results for an event containing only one clearly visible defect (a,b), another containing two visible defects (c,d) and finally a shear banding case (e,f) are displayed. In all cases, the fit corresponds reasonably well, at least visually, in particular further away from the inclusion.In the center of the core equation 4 is not valid, and the displacements can diverge.

\section{Shear Modulus Estimation and error distribution}\label{app:ss}

\begin{figure}[H]
	
	\centering
	\includegraphics[width=.5\linewidth]{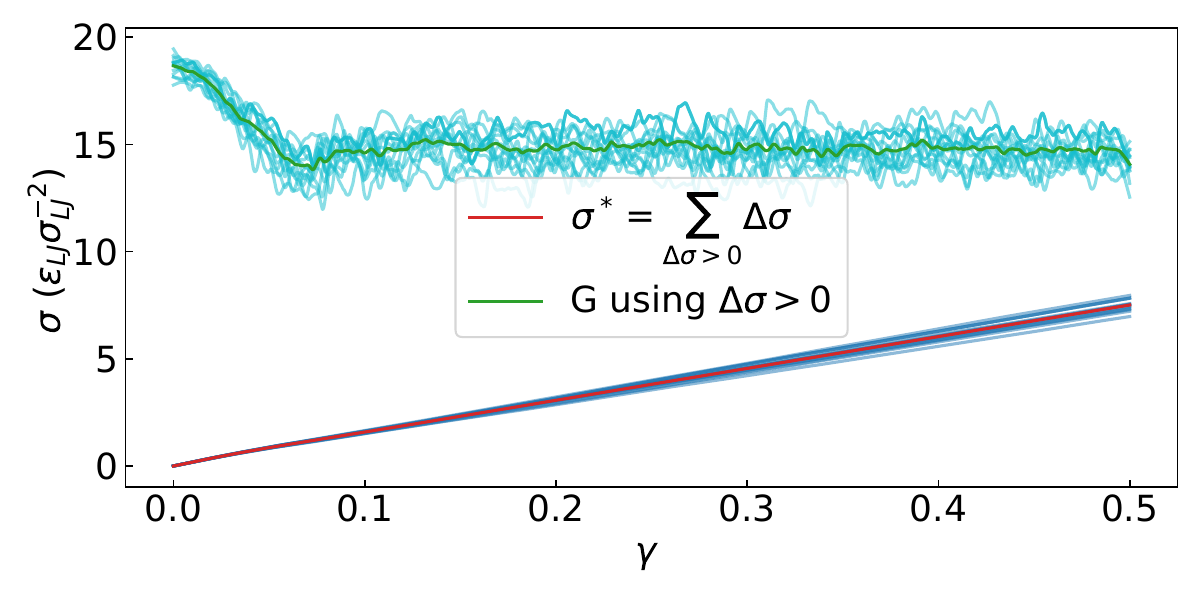}
	\caption{	\label{fig:shearMod}Shear modulus $G$ (green and cyan) obtained from the stress-strain curve (red and blue) with the stress drop removed. The green and red curves are respectively the average modulus and "increase only" stress-strain curves, while the cyan and blue curves give the individual realization for the 14 different repetitions of the process. }
\end{figure}

The shear modulus $G$ is estimated thanks to $G_i=(\sigma_i-\sigma_i^{rev})/\delta\gamma$, $\sigma_i^{rev}$ being the stress in the reverted state of event $i$. In the absence of plastic deformation during the reversion of a single $\delta\gamma$\footnote{Which is true for 99.7 \si{\percent} of the events, and in the case of an outlier, the $G_i$ of the previous event is used.}, this allows us to consider the elastic shear modulus of the inherent state. However, another method can be used: $G_i$ can be obtained from the stress-strain curves. For this, they are expunged of the stress drops to remove the influence of plastic events. These curves are then filtered with a Savitzky–Golay filter, whose result is derived with respect to strain to obtain the shear modulus. The results are displayed in figure \ref{fig:shearMod}. 

\begin{figure}[H]
	
	\centering
	\includegraphics[width=.49\linewidth]{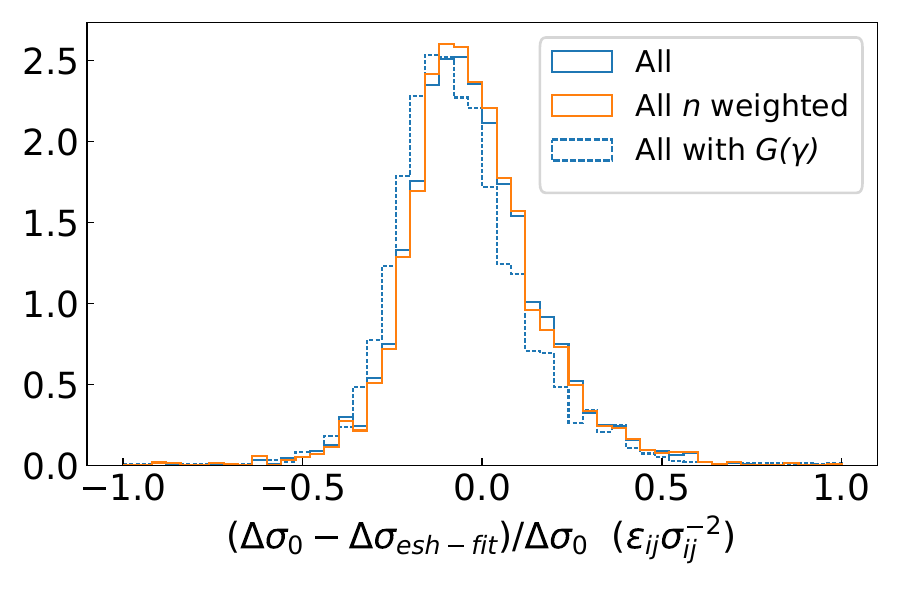}
	
	\caption{	\label{fig:errdist} Normed stress relaxation relative error distribution for events with $\Delta\sigma>$\num{1e-3}, for all events weighted (blue) or not (orange) by the number of events. The full lines correspond to the distribution obtained using the reverted stress estimation of $G$ and the dashed line to the estimation using the stress strain curves.}
\end{figure}

Using the stress-strain curves estimated shear modulus has a relatively small impact on the precision of the estimation of the stress drop. This difference is mostly visible in the shifted distribution of relative error shown in figure \ref{fig:errdist}. This distribution is limited to events with a stress drop superior to $\Delta\sigma>$\num{1e-3} to limit the impact of low stress drop events whose contribution to the global relaxation can be neglected (see figure \ref{fig:1DHist}). Relying on the stress strain curves shifts the relative error from -3 to -8 \si{\percent} and does not impact the variance of the distribution.

\section{ Boundary Conditions and Size Effects }\label{app:se}

In this section, the size effects and periodic boundary conditions are discussed. 
As pointed out in the main article, even though periodic boundary conditions are accounted for in equation 5, equation 4 assumes an infinite medium for a single inclusion \cite{albaret_mapping_2016}. Thus, the model does not account for interactions between inclusion and self-interaction through the periodic boundary conditions. This may induce an error. This effect can be studied through the relative error distribution displayed in figure \ref{fig:errdist}. There is a systematic relative error in the prediction of stress relaxation. However, the distribution remains very similar once weighted by the number of defects considered. This absence of impact of the number of events shows that the underestimation is not due to the interaction between defects. 
This also indicates that the infinite medium approximation does not impact the results at this scale; indeed, it would model self-interaction between inclusion at a distance of a box length when we know that the superposition (sum) of displacement due to multiple defects can give an accurate prediction.

\begin{figure}[H]
	
	\centering
	\includegraphics[width=.49\linewidth]{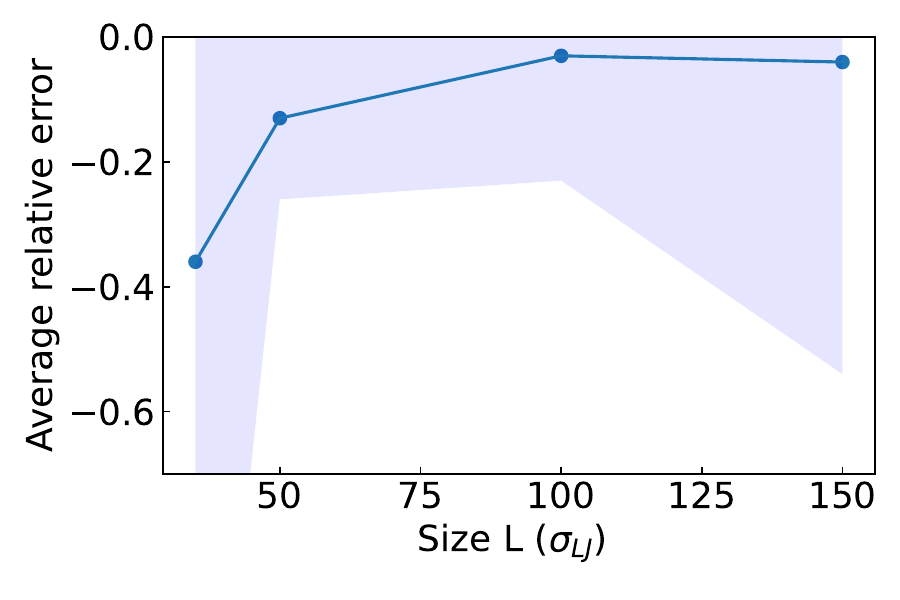}
	
	\caption{	\label{fig:size}Average relative error in stress relaxation for events with a single considered inclusion and $\Delta\sigma>$\num{1e-3} as a function of system size. The shaded area represent the standard deviation.}
\end{figure}

Another method to check for size effects is to study the impact of the size of the simulation supercell. This is provided in figure \ref{fig:size} with the relative error for single events for a few box sizes. For each size, 3 to 14 independent glasses are studied. Although it is not possible to assess the asymptote for large systems, it appears that the error stabilizes around -3\% from a size of 100. This hints that size effects are not the cause of the underestimation of the stresses observed in figure \ref{fig:errdist}.

\bibliographystyle{ieeetr}
\bibliography{FalkGroup}